\documentclass{emulateapj}
\usepackage{graphicx}

\slugcomment{To appear in ApJ}
\shorttitle{Ionized nitrogen at high redshift}
\shortauthors{Decarli et al.}

\def\Lsun{L$_\odot$}
\def\Msun{M$_\odot$}

\def\Hii{H\,{\sc ii}}
\def\Oiii{[O\,{\sc iii}]}

\def\Oi{[O\,{\sc i}]}
\def\Nii{[N\,{\sc ii}]}
\def\nii{[N\,{\sc ii}]$_{\rm 205}$}
\def\niim{[N\,{\sc ii}]$_{\rm 205 \mu m}$}
\def\Cii{[C\,{\sc ii}]}
\def\Ci{[C\,{\sc i}]}

\def\Hii{H\,{\sc ii}}

\def\aco{{\rm CO}(1-0)}
\def\bco{{\rm CO}(2-1)}
\def\cco{{\rm CO}(3-2)}
\def\dco{{\rm CO}(4-3)}

\def\fco{{\rm CO}(6-5)}
\def\gco{{\rm CO}(7-6)}

\def\ico{{\rm CO}(9-8)}
\def\jco{{\rm CO}(10-9)}
\def\kco{{\rm CO}(11-10)}
\def\lco{{\rm CO}(12-11)}
\def\kms{km\,s$^{-1}$}
\def\um{$\mu$m}
\def\mm{MM\,18423+5938}
\def\apm{APM\,08279+5255}
\def\lsim{\mathrel{\rlap{\lower 3pt \hbox{$\sim$}} \raise 2.0pt \hbox{$<$}}}
\def\gsim{\mathrel{\rlap{\lower 3pt \hbox{$\sim$}} \raise 2.0pt \hbox{$>$}}}

\begin{document}

\title{
Ionized nitrogen at high redshift
}

\author{
Decarli R.\altaffilmark{1}, 
Walter F.\altaffilmark{1,2},
Neri R.\altaffilmark{3}, 
Bertoldi F.\altaffilmark{4},
Carilli C.\altaffilmark{5},
Cox P.\altaffilmark{3}, 
Kneib J.P.\altaffilmark{6},
Lestrade J.F.\altaffilmark{7},
Maiolino R.\altaffilmark{8}, 
Omont A.\altaffilmark{9},
Richard J.\altaffilmark{10,11},
Riechers D.\altaffilmark{12},
Thanjavur K.\altaffilmark{13,14},
Weiss A.\altaffilmark{15}
}
\altaffiltext{1}{Max-Planck Institut f\"{u}r Astronomie, K\"{o}nigstuhl 17, D-69117, Heidelberg, Germany. E-mail: {\sf decarli@mpia.de}}
\altaffiltext{2}{NRAO, 520 Edgemont Road, Charlottesville, VA 22903-2475, USA}
\altaffiltext{3}{IRAM, 300 rue de la piscine, F-38406 Saint-Martin d'H\`eres, France}
\altaffiltext{4}{Argelander Institute for Astronomy, University of Bonn, Auf dem H\"{u}gel 71, 53121 Bonn, Germany}
\altaffiltext{5}{NRAO, Pete V.\,Domenici Array Science Center, P.O.\, Box O, Socorro, NM, 87801, USA}
\altaffiltext{6}{Laboratoire d'Astrophysique de Marseille, Observatoire d'Astronomie Marseille-Provence, BP 8, F13376 Marseille, France}
\altaffiltext{7}{Observatoire de Paris, CNRS, 61 Av. de l'Observatoire, F-75014, Paris, France}
\altaffiltext{8}{INAF-Osservatorio Astronomico di Roma, via di Frascati 33, 00040 Monte Porzio Catone, Italy}
\altaffiltext{9}{Institut d'Astrophysique de Paris, UPMC and CNRS, 98bis Bld Arago, F75014, Paris, France}
\altaffiltext{10}{CRAL, Observatoire de Lyon, Universit\'{e} Lyon 1, 9 Avenue Ch. Andr\'{e}, 69561 Saint Genis Laval Cedex, France}
\altaffiltext{11}{Dark Cosmology Centre, Niels Bohr Institute, University of Copenhagen, Juliane Maries Vej 30, DK-2100 Copenhagen, Denmark}
\altaffiltext{12}{Astronomy Department, Caltech, 1200 East California boulevard, Pasadena, CA 91125, USA}
\altaffiltext{13}{Canada France Hawaii Telescope Corporation, HI 96743, USA}
\altaffiltext{14}{Department of Physics \& Astronomy, University of Victoria, Victoria, BC, V8P 1A1, Canada}
\altaffiltext{15}{Max-Planck-Institut f\"ur Radioastronomie, Auf dem H\"ugel 69, 53121 Bonn, Germany}

\begin{abstract}
We present secure \niim{} detections in two mm-bright, strongly 
lensed objects at high redshift, \apm{} ($z$=3.911) and \mm{} ($z$=3.930), 
using the IRAM Plateau de Bure Interferometer. Due to its ionization energy 
\niim{} is a good tracer of the ionized gas phase in the interstellar medium.
The measured fluxes are $S$(\niim)=$(4.8\pm0.8)$ Jy\,\kms{} and $(7.4\pm0.5)$
Jy\,\kms{} respectively, yielding line luminosities of 
$L($\nii$) =(1.8\pm0.3)\times10^{9} \,  \mu^{-1}$ \Lsun{} for \apm{} and 
$L($\nii$) =(2.8\pm0.2)\times10^{9} \, \mu^{-1}$ \Lsun{} for \mm{}. Our 
high-resolution map of the \niim{} and 1 mm continuum emission in \mm{} 
clearly resolves an Einstein ring in this source, and reveals a velocity 
gradient in the dynamics of the ionized gas. A comparison of these maps
with high-resolution EVLA CO observations enables us to perform the first 
spatially-resolved study of the dust continuum-to-molecular gas surface brightness
($\Sigma_{\rm FIR}\propto\Sigma_{\rm CO}^N$, which can be interpreted as the 
star formation law) in a high-redshift object. We find a steep relation
($N=1.4\pm0.2$), consistent with a starbursting environment.
We measure a \niim{}/FIR 
luminosity ratio in \apm{} and \mm{} of 9.0 $\times 10^{-6}$ and 5.8 
$\times 10^{-6}$, respectively. This is in agreement with the decrease of 
the \niim{}/FIR ratio at high FIR luminosities observed in local galaxies. 
\end{abstract}

\section{Introduction}

Forbidden atomic fine structure transitions are important cooling
lines of the interstellar medium (ISM). They provide effective cooling
in cold regions where allowed atomic transitions can not be excited,
and thus are critical diagnostic tools to study the star-forming
ISM.  Perhaps the most important cooling line is the forbidden
$^2$P$_{3/2}\rightarrow^2$P$_{1/2}$ fine-structure line of ionized
carbon (\Cii) at 158 \um, which alone accounts for 0.1 -- 1 percent of
the total continuum far-infrared (FIR) luminosity in local, star-forming
galaxies \citep[see, e.g.,][]{malhotra01}. Other main cooling atomic 
lines are the oxygen \Oi{} (63 \um{} and 145 \um) and \Oiii{} (52 \um{} 
and 88 \um) lines, as well as the nitrogen \Nii{} lines at 122 \um{} and 205 
\um.  

As the ionization potential of carbon is 11.3\,eV (hydrogen: 13.6\,eV), 
\Cii{} is a tracer for both the neutral atomic and ionized medium, 
predominantly of photon-dominated regions. The ionization potentials for 
oxygen and nitrogen, on the other hand, are 13.6\,eV and 14.5\,eV, 
respectively, implying that their ions trace of the ionized medium.  The 
\niim{} transition is of particular interest as it has
a critical density in the ionized medium that is very close to that of 
\Cii{}, thus potentially 
providing complementary information on the origin of the \Cii{} emission 
\citep[e.g.,][]{oberst06,walter09a}.

\begin{figure}[h]
\includegraphics[width=0.99\columnwidth]{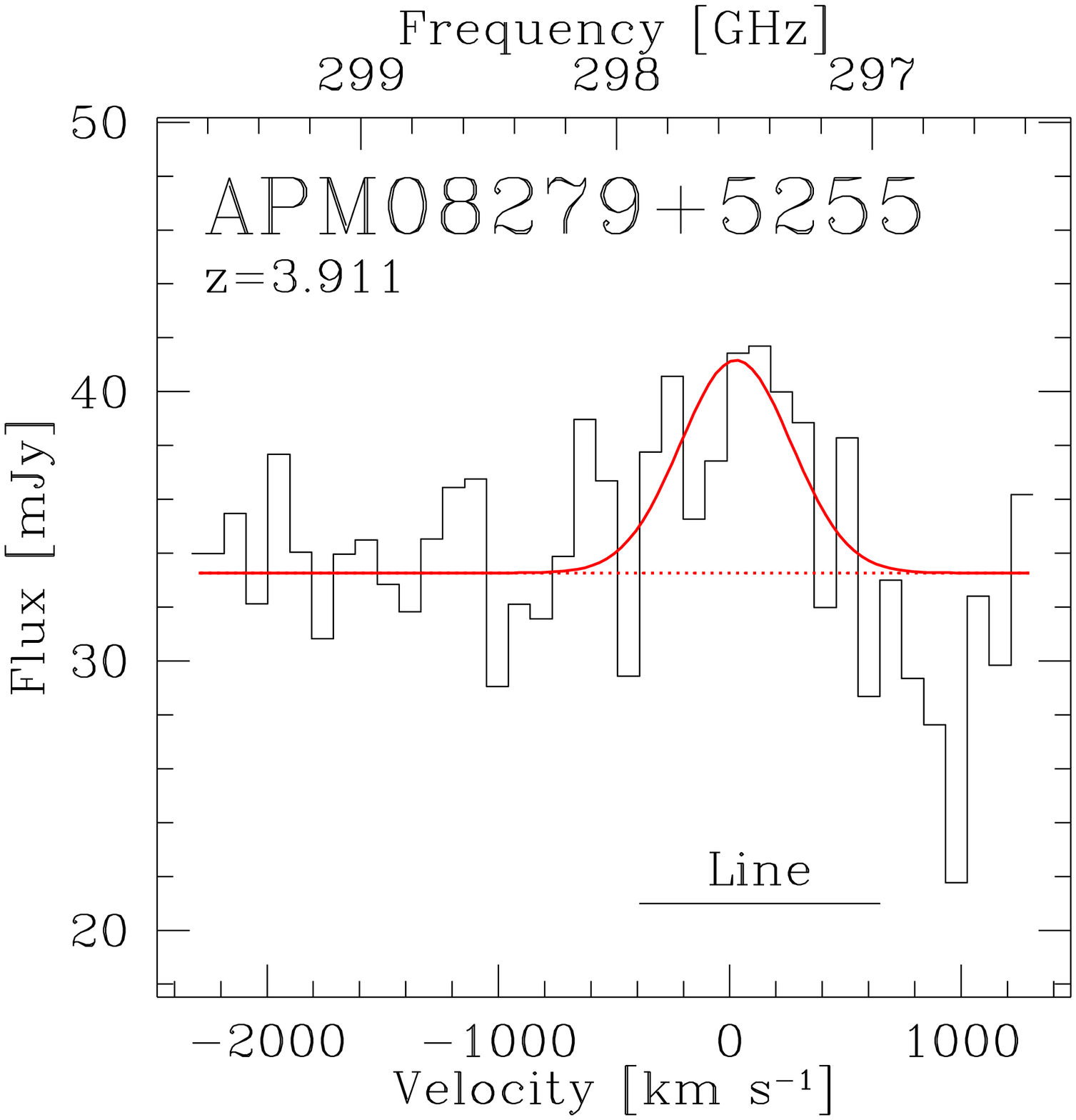}\\
\includegraphics[width=0.94\columnwidth]{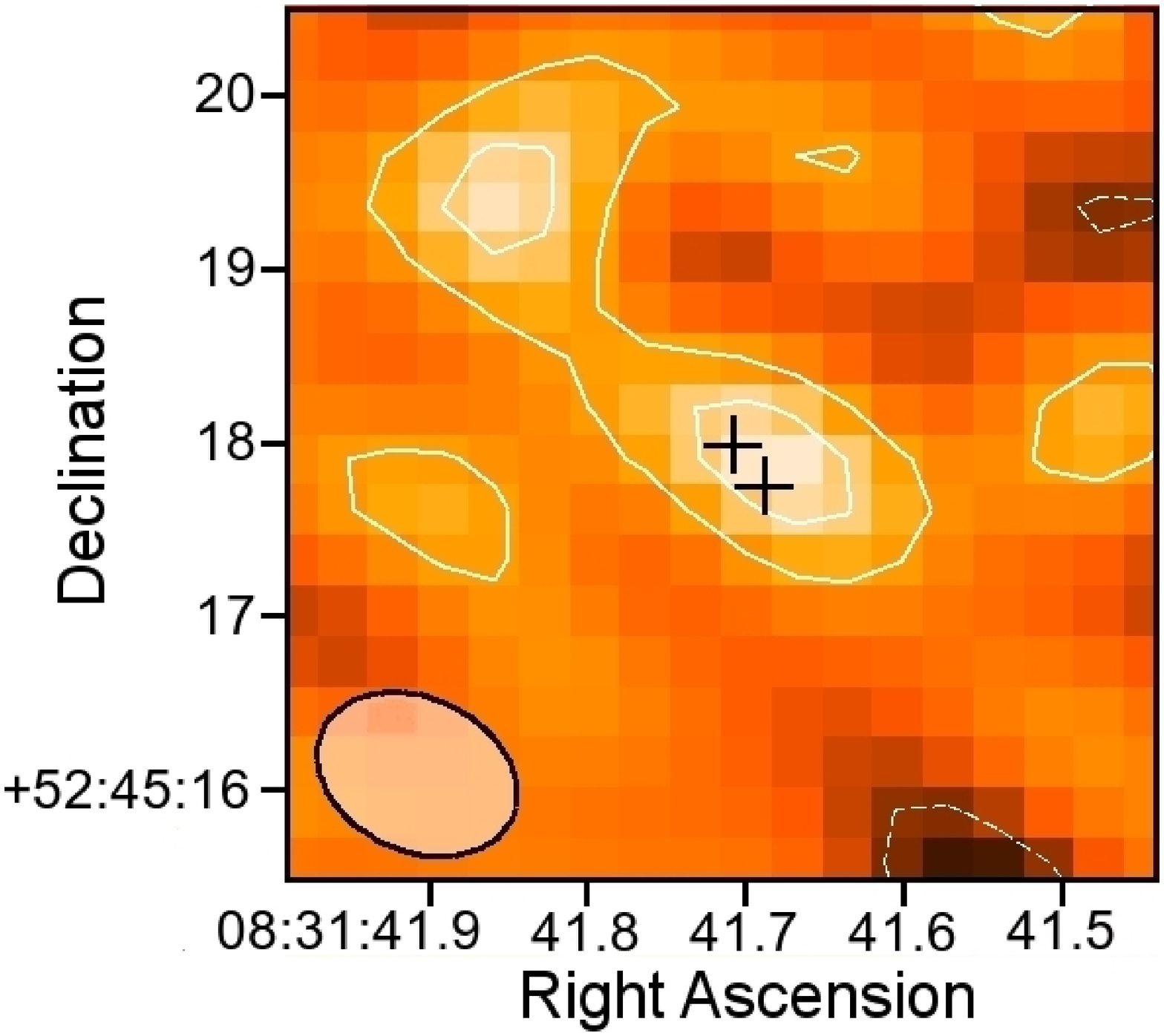}\\
\caption{Observed spectrum ({\it top}) and continuum-subtracted map 
({\it bottom}) of the \niim{} emission in \apm{}. The spectrum is resampled
into 95 \kms{} wide bins. The velocity scale is set assuming the CO 
redshift, $z$=$3.911$. The continuum and line fits are shown with 
thick, solid red lines. The channels used to create the \niim{} line map are 
marked by a horizontal line. In the map, solid thick (dashed thin) contours are
positive (negative) isophotes, separated by 2-$\sigma$ (1-$\sigma$=0.47 mJy 
beam$^{-1}$). The synthetic beam ($1.2''\times0.9''$) is shown in the 
bottom-left corner. 
The two crosses mark the position of the two lensed images reported by 
\citet{riechers09}.}
\label{fig_apm}
\end{figure}

\begin{figure}
\includegraphics[width=0.99\columnwidth]{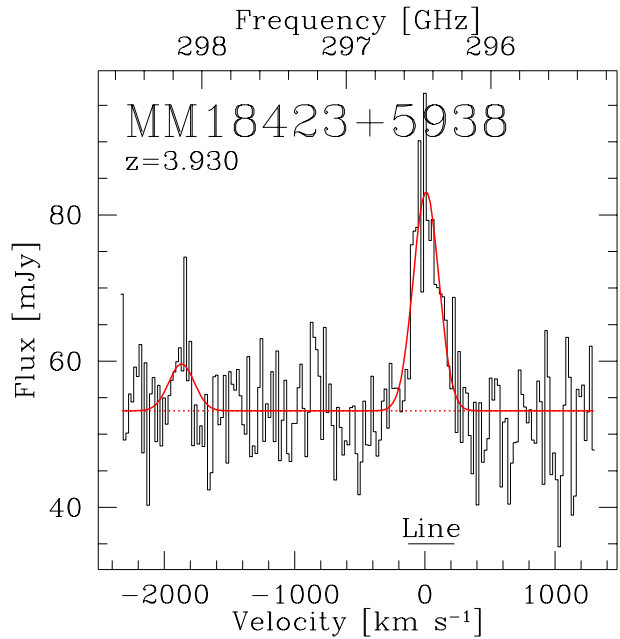}\\
\caption{Integrated spectrum of the \niim{} plus continuum emission in 
\mm{}. Data are 
resampled into 20 \kms{} wide bins. The continuum and line fits are shown 
with thick, solid red lines. The tentative SO$_2$ detection at 298.23 GHz is
also fitted by forcing the same line width as observed for \niim{}. The 
channels used to create the \niim{} line map are marked by a horizontal line.}
\label{fig_mm_spc}
\end{figure}

\begin{figure*}
\includegraphics[width=0.99\textwidth]{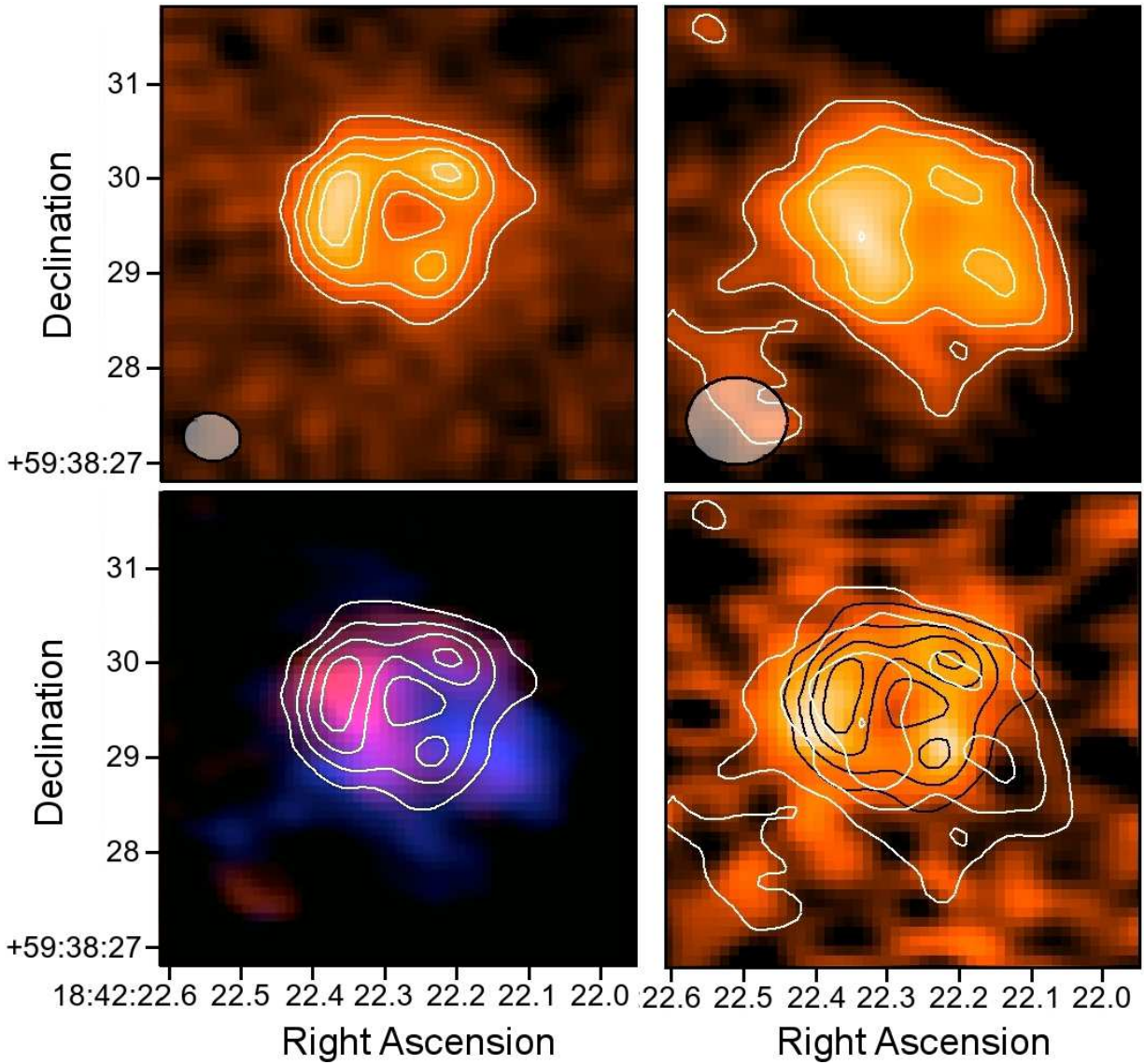}\\
\caption{{\it Top left:} High-resolution (i.e., uniform-weighted) map of 
the pure continuum emission in \mm{}. The Einstein ring is clearly resolved.
Contours show the +4, +8, +12, +16-$\sigma$ isophotes (1-$\sigma$=0.58 
mJy\,beam$^{-1}$). The synthetic beam ($0.6''\times0.5''$) is also shown in 
the bottom-left corner. {\it Top right:} 
Continuum-subtracted map of the \niim{} emission. Natural weighting is 
used here (beam size: $1.0''\times0.9''$). Contours are separated by 2-$\sigma$ (1-$\sigma$=0.82 
mJy\,beam$^{-1}$). {\it Bottom left:} Continuum-subtracted maps of the red 
and blue wings of the \niim{} emission in \mm. The two sides of the line 
peak in opposite sides of the Einstein ring, unveiling a velocity gradient 
in the source. The contours from the continuum map in the top-right panel 
are also shown for reference. {\it Bottom right:} Comparison among the 
\niim{} line map (white contours), the continuum emission at 1 mm 
(black contours) and the \bco{} map published by \citet{lestrade11} (color 
scale). The continuum emission is brighter in the Eastern and 
Northern sides of the ring, while CO and \niim{} emission peaks in the East 
and South-West.}
\label{fig_mm_all}
\end{figure*}

\begin{figure*}[t]
\includegraphics[width=0.49\textwidth]{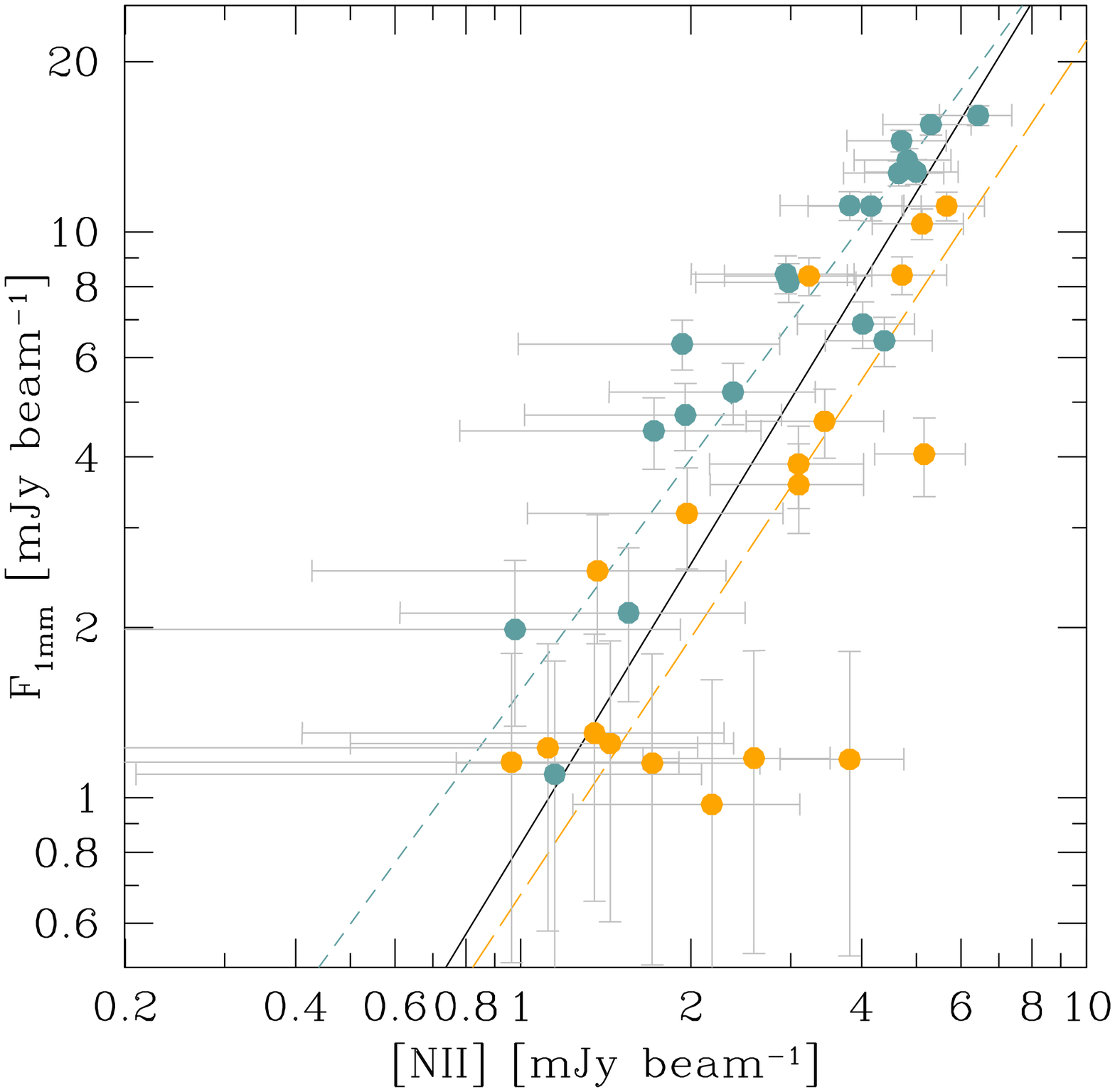}
\includegraphics[width=0.49\textwidth]{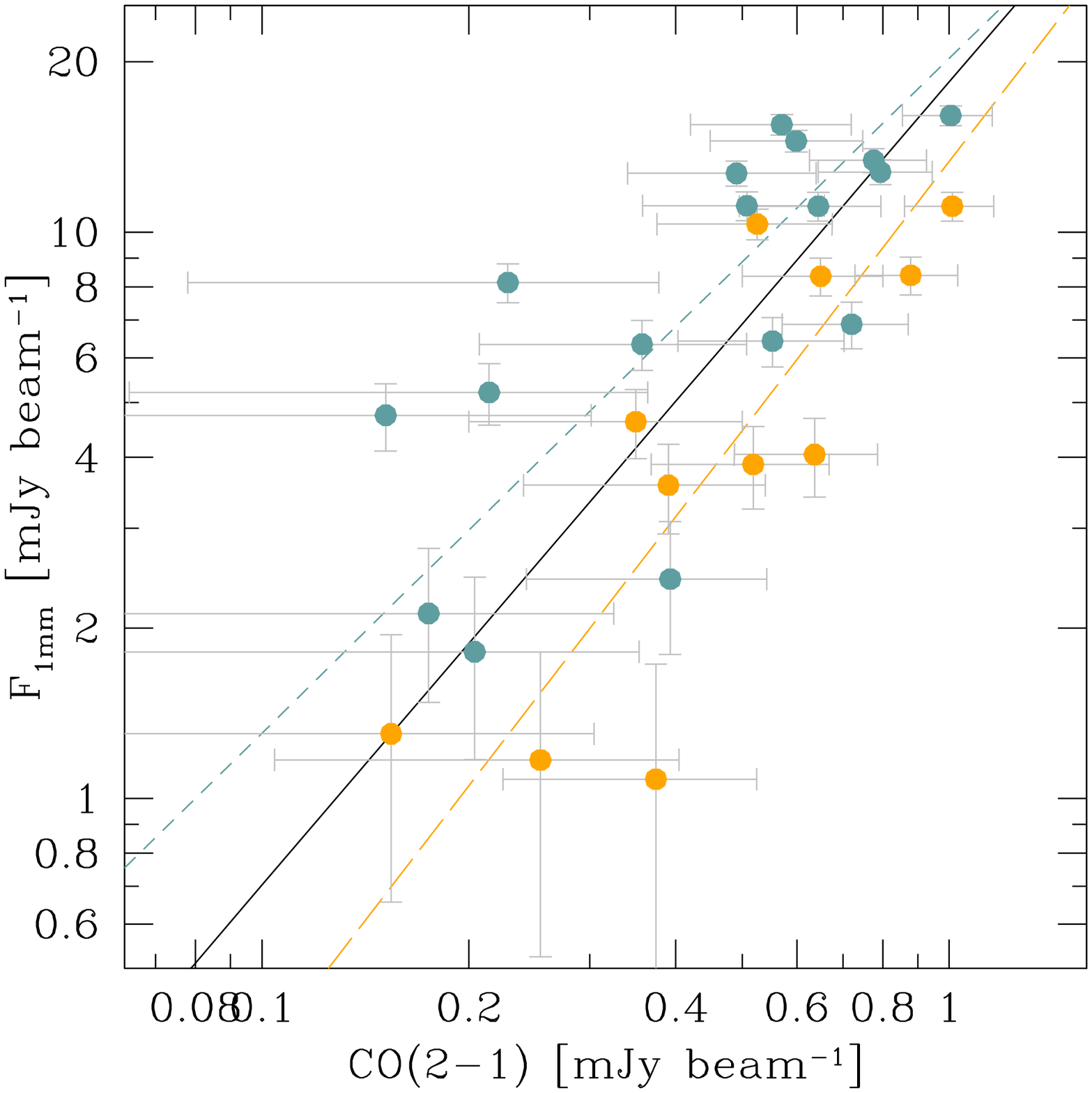}\\
\caption{Comparison between the FIR continuum surface brightness at 1 mm 
(observed frame) and the surface brightness of \niim{} ({\em left panel}) 
and \bco{} ({\em right panel}) in our PdBI and EVLA observations of \mm{}. 
Since the FIR surface brightness can be used as a proxy for the surface 
density of star formation ($\Sigma_{\rm SFR}$), while the CO surface 
brightness maps the surface density of molecular gas ($\Sigma_{\rm H_2}$), 
the right panel shows the first, spatially-resolved star-formation law
in a high-$z$ galaxy. Each point is a $0.5''\times0.5''$ pixel with 
$>$1-$\sigma$ flux. Error bars show the corresponding 1-$\sigma$ 
uncertainties. Grey (orange) points refer to the Northern (Southern) part 
of the Einstein ring, with a cut at declination=+59:38:29.3. The FIR 
emission increases with the \niim{} and \bco{} emission 
($\Sigma_{\rm F_{\rm 1mm}}\propto\Sigma_{\rm [NII]}^{1.4}$ and
$\Sigma_{\rm F_{\rm 1mm}}\propto\Sigma_{\rm CO}^{1.6}$); best fits on the total,
Northern and Southern pixels are shown as black solid, grey short-dashed
and orange long-dashed lines respectively. The FIR continuum is $\sim2$ 
times brighter at any given line luminosity in the 
Northern part of the ring with respect to the Southern part, compared
to both \niim{} and \bco{}.
}
\label{fig_sk}
\end{figure*}

\begin{figure}[h]
\includegraphics[width=0.49\textwidth]{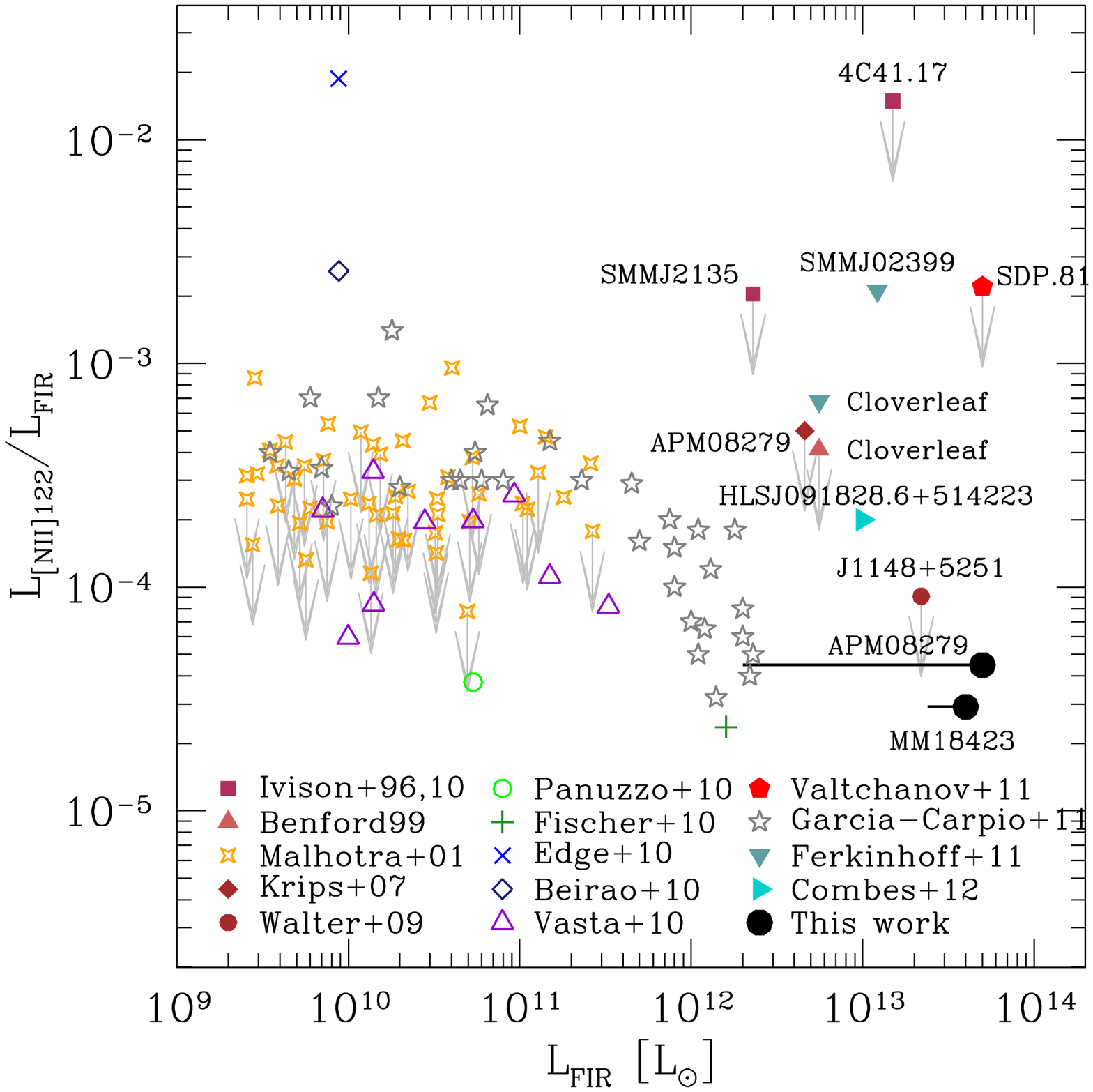}\\
\caption{\Nii$_{122 \mu{\rm m}}$/FIR luminosity ratio, as a function of the 
FIR continuum luminosity. When the 122 \um{} line is not available
(like in the sources presented here), we assumed a \Nii{} 122-to-205 \um{} 
luminosity ratio of 5 \citep[see, e.g.,][]{beirao10}. High redshift sources 
(filled symbols) are labeled. The horizontal bars mark the uncertainties in 
the magnification correction of our sources (4--100 for \apm{}, see
Weiss et al. 2007; Riechers et al. 2009; 12--20 for \mm{}, see Lestrade 
et al. 2011).
}
\label{fig_nii_fir}
\end{figure}

In studies of star formation at high-$z$, the need for diagnostics other
than \Cii{} is also motivated by two other reasons: 1) The relative 
intensities of \Cii{} and \niim{} is susceptible to N/C abundance variations 
\citep[e.g.,][]{matteucci93}, thus acting as diagnostics of metal 
enrichment in the first galaxies. 2) The \Cii{} line falls out of the 1.3mm
window at $z\approx8$, and will not be observable at $z$ up to 10.2 with 
ALMA, as no 2mm receiver (ALMA `band 5') will be available for a large 
number of antennas. This limits \Cii{} studies in the first 
galaxies observable at the beginning of cosmic reionization.

Whereas the \Cii{} line has now been abundantly detected in the local
universe \citep[e.g.][]{stacey91,madden97,luhman98,malhotra01,beirao10,
edge10,fischer10,ivison10,loenen10}, and is
now (almost) routinely detected at high redshift 
\citep[][]{maiolino05,maiolino09,iono06,walter09b,wagg10,stacey10,cox11,debreuck11}, 
measurements of the \niim{} line are scarce. The \niim{} line was first 
detected by {\em FIRAS} aboard {\em COBE} in the Milky Way 
\citep{wright91}, and later in the Galactic H\,{\sc ii} regions G333.6--0.2 
\cite{colgan93} and DR21 \citep{white10} and in the Carina Nebula 
\citep{oberst06}. The \niim{} line has also been recently detected in some local 
galaxies, e.g., NGC 1097 \citep[Beir\~{a}o et al. 2010; see also][]{garcia11}.
On the other hand, little is known about nitrogen at high redshift. 
Multiple attempts were performed to detect this line at high redshift 
(4C41.17 and PC2047+0123: Ivison \& Harris 1996, Cloverleaf: Benford 1999, 
APM\,08279: Krips et al. 2007, J1148+5251: Walter et al.\ 2009b; 
SDP.81: Valtchanov et al. 2011). However, all these measurements,
but the one on the $z$=$6.42$ quasar J1148+5251, were too insensitive by
about an order of magnitude to potentially detect the \niim{}
line. Very recently, \citet{ferkinhoff11} reported the first detection of 
the second line of ionized nitrogen, \Nii{}$_{\rm 122 \mu m}$ in two $z\sim2.7$ targets,
SMMJ02399-0136 ($z$=$2.81$) and the Cloverleaf QSO ($z$=$2.56$). 
\citet{bradford11} published a tentative (1.5-$\sigma$) detection of \niim{} 
in the lensed quasar \apm{} ($z$=$3.911$), and \citet{combes12} reported the 
detection of the \niim{} line in a lensed submillimeter galaxy, 
HLSJ091828.6+514223, at $z=5.2$.

Here we present secure detections of the \niim{} fine-structure
line in two high-$z$ sources, the lensed quasar \apm{} ($z$=$3.911$) 
and the sub-millimeter galaxy (SMG) \mm{} ($z$=$3.930$). These sources are 
strongly magnified by gravitational lensing (magnification factor $\mu$=4--100
for \apm{} and 12--20 for \mm{}; see Egami et al. 2000, Riechers et al. 2009,
Lestrade et al. 2011), and represent two of the brightest molecular emitters 
at this redshift, with \fco{} peak flux densities of 7.3\,mJy \citep{weiss07} 
and 33\,mJy \citep{lestrade10}, respectively.  Observations were
carried out at the IRAM Plateau de Bure Interferometer (PdBI), and are summarized 
in Sec.\,\ref{sec_obs}. Results are presented in Sec.\,\ref{sec_results}.

Throughout the paper we will assume a standard cosmology with $H_0=70$ km 
s$^{-1}$ Mpc$^{-1}$, $\Omega_{\rm m}=0.3$ and $\Omega_{\Lambda}=0.7$.

\section{Observations}\label{sec_obs}

In our observations we exploited the capabilities of the new `band 4' 
receiver at PdBI. Covering the frequency range between 277 and 371 GHz (with
a small gap due to atmospheric absorption at rougly 320--330 GHz), this band
opens up the opportunity to search for the \niim{} line in a wide redshift 
range ($2.85<z<4.27$).

\apm{} was observed in compact array configuration with 5 antennas
(6Cq-E10) on March 8th, 2011. Baselines ranged between 20 and 140 m.
3C273, 3C84 and MWC349 were used as amplitude calibrators. The quasar 
0917+624 was observed every 30 min for phase calibration. The time on
sources was 4.5 hours (5-antenna equivalent). \mm{} was observed in 
both compact and extended configurations (6Cq-E10 and 6Bq) between 
January 3rd and March 9th, 2011. Baselines ranged between 17 and 446 
m. 3C273, 3C345, MWC349 and 3C84 were used as amplitude calibrators, while 
1849+670 was observed as phase calibrator. The total time on source 
was 10 hours (6-antenna equivalent).

The tuning frequencies were 297.522 and 296.400 GHz respectively, based on 
the CO redshift of the sources \citep{weiss07,lestrade10,lestrade11}. The 
receiver worked in Lower-Side Band. System temperature ranged between 150 
and 350 K. Data reduction and analysis was performed using the most recent 
version of the GILDAS package. Maps were extracted using natural weighting. 
This allows us to fully recover the flux of our sources, given that 
their spatial extent ($\lsim2''$) is comparable with or smaller than the 
angular scale filtered in by the smallest baselines in our observations 
($\sim2''$). 
The resulting synthetized beams are $1.2''\times0.9''$ for \apm{} and
$1.0''\times0.9''$ for \mm{}. In order to take full advantage of the
high spatial resolution observations of \mm{}, we also extracted a 
pure continuum map of this source using uniform weighting. This high
resolution map has a synthetic beam size of $0.6''\times0.5''$, at the
price of filtering out a significant fraction ($\sim66$\%) of the flux
from the extended emission of the object. Therefore, in the remainder of 
the analysis, all the flux measurements will refer only to the map 
obtained with natural weighting. The 1-$\sigma$ noise per 20 MHz-wide 
channel ($\approx$ 20.2 \kms{}) is 3.4 mJy for \apm{} and 2.4 mJy for \mm{},
corresponding to a 1-$\sigma$ sensitivity of $0.66$ mJy\,beam$^{-1}$ and
$0.82$ mJy\,beam$^{-1}$ over the line width (see Sec.\,\ref{sec_results}).

\section{Results}\label{sec_results}

\subsection{\niim{} and continuum emission}

\subsubsection*{\apm}

Figure \ref{fig_apm} shows the observed spectrum and the 
(continuum-subtracted) line map of \apm{}. The \niim{} line is detected
at modest significance. We fitted the spectrum with a flat continuum plus a 
gaussian profile for the \niim{} emission. The fitted \niim{} flux is 
$S$(\niim)=$4.8\pm0.8$ Jy\,\kms{}, consistent with the tentative detection 
($7.6\pm5.4$ Jy\,\kms{}) reported in \citet{bradford11}. We measure a line width of 
$570\pm110$ \kms{}, consistent with the weighted average of the line width 
values from CO transitions reported in \citet{weiss07}.
Line luminosities $L_c$ and $L'$ are derived following \citet{solomon92},
and reported in Table 1, together with all the relevant numbers and fitted 
parameters. We measure a continuum flux of $33\pm3$ mJy, in agreement with 
the extrapolation between the SCUBA and PdBI observations at 850 \um{} and 
1.4mm respectively \citep{weiss07}. 

\subsubsection*{\mm}

In Figure \ref{fig_mm_spc} we show the integrated spectrum of the \niim{}
emission in \mm{}. The line is detected at very high significance. The 
gaussian fit gives an integrated \niim{} flux of $S$(\niim)=$7.4\pm0.5$ 
Jy\,\kms{} and a line width of $230\pm20$ \kms{} (for a comparison, the 
\aco{} and \bco{} lines reported in Lestrade et al. 2011 have widths of 
$160\pm30$ and $240\pm30$ \kms{} respectively). Another line is 
tentatively detected at $298.23\pm0.02$ GHz, i.e., at rest frequency 
$1470.22\pm0.11$ GHz, consistent with three different transitions of the 
Sulfur dioxide, SO$_2$ (at 1470.225, 1470.327 and 1470.342 GHz 
respectively). Given the low S/N of this line ($\sim3.5$-$\sigma$), we 
fitted it with a Gaussian by imposing the same line width as observed 
for \niim{}. We find a continuum flux (integrated over the spatial
extension of the emission) of $53\pm2$ mJy, consistent with the 1.2 mm
MAMBO flux reported in \citet{lestrade10}, assuming a grey-body dust with 
opacity index $\beta$=1.

Our PdBI observations clearly resolved the 1mm continuum emission of \mm{}.
The Einstein ring reported in \citet{lestrade11} is clearly seen in the
pure-continuum, high resolution map shown in Figure \ref{fig_mm_all} ({\em
top-left panel}). Also the \niim{} emission appears clearly extended
even in the lower resolution maps shown in the top-right panel of
Figure \ref{fig_mm_all}. The \niim{} line emission shows a clear velocity
gradient from North-East (red-shifted) to South-West (blue-shifted).
This is highlighted in Figure \ref{fig_mm_all} ({\em bottom-left panel})
where we overplot the maps of the red and blue wings of the \niim{} line. We 
measure peak-to-peak velocity differences of $\sim 180$ \kms{} and a
velocity dispersion $\sim 80$ \kms{}. The $v/\sigma>1$ value indicates 
ordered kinematics for the ionized gas in the lensed source.
A detailed model of the lens is required to reconstruct the intrinsic 
brightness of the source from these images and to properly constrain the 
dynamics of the system. This is beyond the scope of this paper.

\subsection{\niim{} and CO}

Here we compare our \niim{} observations with the available literature
data on CO emission in the two targets. In particular, the \niim{} to 
\fco{} luminosity ratio in our sources (\niim{}/\fco{}=$1.5\pm0.4$ in 
\apm{} and $2.3\pm0.3$ in \mm{}) is similar to the values reported 
by \citet{white10} in the Galactic region DR21 (\niim{}/\fco{}=$1.26\pm0.35$),
by \citet{panuzzo10} in M82 (\niim{}/\fco{}=$1.68 \pm 0.05$)
and by \citet{vanderwerf10} in Mrk231 (\niim{}/\fco{}$\sim1.2$). 

In Figure \ref{fig_mm_all} ({\em bottom-right panel}) we compare our \niim{} 
map of \mm{} with our map of the 1mm continuum emission and the \bco{}
observations presented in \citet{lestrade11}, which have similar S/N and 
spatial resolution as the \niim{} data presented here. Since nitrogen 
emission is tracing the ionized gas, while the CO emission maps 
the molecular gas (which is the fuel of star formation) and the FIR 
continuum traces the distribution of dust heated by young stars,
this comparison allows for a direct, spatially-resolved study of three 
important components of the ISM that are related to star formation. From 
the Figure, it is apparent that
the emission from the three tracers shows different morphologies (e.g., 
most of \niim{} and \bco{} emission arises in two blobs in the East and 
South-West parts of the Einstein ring, while a bright continuum emission is 
observed also in the Northern side of the ring). 

More quantitatively, in Figure \ref{fig_sk} we perform a pixel-by-pixel 
comparison of the emission of the FIR continuum with respect to the 
\niim{} and \bco{} lines\footnote{A continuum map obtained with natural
weighting is used here, so that the beam size of the three maps (FIR 
continuum, \niim{} and \bco{}) are similar.}. In order to avoid 
over-sampling, we consistently rebinned all the maps into $0.5''\times0.5''$
pixels, and considered only pixels with $>$1-$\sigma$
flux in each axis. We find that the surface brightness of
the continuum, $\Sigma_{\rm FIR}$, shows a steep correlation with the 
surface brightnesses of \niim{} and \bco{}, $\Sigma_{\rm [NII]}$ and 
$\Sigma_{\rm CO}$: $\Sigma_{\rm FIR}\propto\Sigma_{\rm [NII]}^{1.6\pm0.1}$ and
$\Sigma_{\rm FIR}\propto\Sigma_{\rm CO}^{1.4\pm0.2}$. The latter relation
is of particular interest, as $\Sigma_{\rm FIR}$ can be used as a proxy for
the star formation surface density, $\Sigma_{\rm SFR}$ \citep{kennicutt98}, 
while $\Sigma_{\rm CO}$ traces the surface density of the molecular gas, 
$\Sigma_{\rm H_2}$, which is the fuel for star formation. The right-hand
panel of Figure \ref{fig_sk} represents therefore the first 
spatially-resolved study of the star formation law in a high-$z$ galaxy. 

Locally, star formation surface density scales, to first order, linearly 
with the molecular gas surface density 
\citep[$\Sigma_{\rm SFR}\propto\Sigma_{\rm H_2}^{N}$, with $N\approx1$; 
see, e.g.,][]{bigiel08,leroy08,schruba11}. The relation steepens if 
one considers high-density environments and molecule-rich galaxies 
\citep[e.g.,][]{kennicutt98,wong02,daddi10,genzel10}. This appears to be the
case of \mm{}, where we measure a slope significantly larger than 1 
($N=1.4\pm0.2$). The steep relation observed between $\Sigma_{\rm FIR}$
and $\Sigma_{\rm [NII]}$ may indicate that, as SFR increases, the ionization 
state of nitrogen changes, with an increasing fraction of multiply ionized
N in the regions of most intense star formation.
Moreover, if we divide the Einstein ring of \mm{} in two
parts, North and South, with a cut at declination +59:38:29.3
(roughly corresponding to the center of the ring), we find that each part
of the ring follows different power-laws. The Northern part of the ring
has a $\sim2$ times brighter continuum for a given \niim{} or \bco{} 
emission. The $\Sigma_{\rm FIR}\propto\Sigma_{\rm CO}^{N}$ relation
shows a marginally flatter slope ($N$=1.2) in the Northern part than
in the Southern part ($N$=1.6), in agreement with the relatively higher 
molecular content in the Southern part of the ring.

\begin{deluxetable*}{lcccc}
\tablecaption{Line and continuum properties in \apm{} and \mm{}} 

\tablehead{\colhead{Quantity} & \colhead{Units} & \colhead{\apm{}} & \colhead{\mm}  & \colhead{References}} 
\startdata
  Redshift		    &	    & 3.911			 & 3.930     & 1,2	      \\
  $D_{\rm L}$		    & [Gpc] & 34.897			 & 35.097    &  	      \\
  $\mu$ 		    &	    & 4--100			 & 12--20    & 3,4,2	      \\
  $\nu_{\rm obs}$           & [GHz] & 297.522                    & 296.400   & 0              \\
  1-$\sigma$ RMS (20 MHz)   & [mJy\,beam$^{-1}$] & 3.4            & 2.4       & 0              \\
\hline
\vspace{-0.8mm}\\
\multicolumn{5}{c}{\em \nii} \\
\vspace{-1mm}\\
  $S$(\niim{})		      & [Jy \kms]			     & $4.8\pm0.8$  & $7.4\pm0.5$  & 0 \\
  FWHM  		      & [\kms]  			     & $570\pm110$  & $230\pm20$   & 0 \\
  $L_{\rm c}$(\niim{})	      & [$10^9 \mu^{-1}$ \Lsun] 	     & $1.8\pm0.3$  & $2.8\pm0.2$  & 0 \\
  $L'$(\niim{})  	      & [$10^{10} \mu^{-1}$ K \kms{} pc$^2$] & $1.81\pm0.3$  & $2.8\pm0.2$ & 0 \\
  $L_{\rm c}$(\niim{})/$L_{\rm FIR}$   & [$10^{-6}$]			     & $9.0\pm1.5$  & $5.8\pm0.4$  & 0 \\
  $M$(H\,{\sc ii})            & [$10^9 \mu^{-1}$ \Msun] 	     & $\gsim4.1$   & $\gsim6.4$   & 0 \\
  $M$(H\,{\sc ii})/$M$(H$_2$) &                      		     & $\gsim0.8$\% & $\gsim2.9$\% & 0 \\
\hline
\vspace{-0.8mm}\\
\multicolumn{5}{c}{\em Other lines} \\
\vspace{-1mm}\\
  $L_{\rm c}$(\Ci{}$_{1-0}$) & [$10^9 \mu^{-1}$ \Lsun]  & $0.118  \pm  0.016$ & $0.29	\pm   0.06 $ & 5,2 \\
  $L_{\rm c}$(\Ci{}$_{2-1}$) & [$10^9 \mu^{-1}$ \Lsun]  & $<0.23	    $ & $0.88	\pm   0.17 $ & 6,2 \\
  $L_{\rm c}$(\aco{})	     & [$10^9 \mu^{-1}$ \Lsun]  & $0.0050 \pm 0.0004$ & $0.0129 \pm 0.0017 $ & 3,7 \\
  $L_{\rm c}$(\bco{})	     & [$10^9 \mu^{-1}$ \Lsun]  & $0.048  \pm  0.011$ & $0.179  \pm  0.018 $ & 3,7 \\
  $L_{\rm c}$(\cco{})	     & [$10^9 \mu^{-1}$ \Lsun]  & $0.196  \pm  0.018$ &  ---		     & 8   \\
  $L_{\rm c}$(\dco{})	     & [$10^9 \mu^{-1}$ \Lsun]  & $0.44   \pm	0.02$ & $0.59	\pm   0.06 $ & 1,2 \\
  $L_{\rm c}$(\fco{})	     & [$10^9 \mu^{-1}$ \Lsun]  & $1.2    \pm	 0.2$ & $1.19	\pm   0.11 $ & 1,2 \\
  $L_{\rm c}$(\gco{})	     & [$10^9 \mu^{-1}$ \Lsun]  &   --- 	      & $0.82	\pm   0.10 $ &   2 \\
  $L_{\rm c}$(\ico{})	     & [$10^9 \mu^{-1}$ \Lsun]  & $3.16   \pm	0.16$ &  ---		     & 1   \\
  $L_{\rm c}$(\jco{})	     & [$10^9 \mu^{-1}$ \Lsun]  & $3.5    \pm	 0.6$ &  ---		     & 1   \\
  $L_{\rm c}$(\kco{})	     & [$10^9 \mu^{-1}$ \Lsun]  & $3.7    \pm	 0.6$ &  ---		     & 1   \\
  $L_{\rm c}$(\lco{})	     & [$10^9 \mu^{-1}$ \Lsun]  & $3.5    \pm	 1.8$ &  ---		     & 9   \\
  $L_{\rm c}$(SO$_2$)	     & [$10^9 \mu^{-1}$ \Lsun]  &   --- 	      & $0.60   \pm   0.16 $ & 0   \\
  $M$(H$_2$)	             & [$10^{11} \mu^{-1}$ \Msun] & $5.3 \pm     0.5$ & $2.2    \pm    0.3$  & 3,7 \\
\hline
\vspace{-0.8mm}\\
\multicolumn{5}{c}{\em Continuum}\\
\vspace{-1mm}\\
  $S$(850\micron)     & [mJy]			   & $75\pm4$	      & -- 	        & 1   \\
  $S$(1mm)   	      & [mJy]			   & $33\pm3$	      & $53\pm2$	&   0 \\
  $S$(1.3mm) 	      & [mJy]			   & $16.9\pm2.5$     & $30\pm2$	& 1,2 \\
  $L_{\rm FIR}$       & [$10^{13} \mu^{-1}$ \Lsun] & $20 $ 	      & $48$		& 1,2 \\
\enddata
\tablerefs{0: This work. 
1: \citet{weiss07}. 
2: \citet{lestrade10}. 
3: \citet{riechers09}. 
4: \citet{egami00}.
5: \citet{wagg06}. 
6: \citet{walter11b}.
7: \citet{lestrade11}. 
8: \citet{downes99}. 
9: \citet{bradford11}.
}
\end{deluxetable*}

\subsection{Ionized and molecular gas masses}

Following \citet{ferkinhoff11}, we can compute the minimum mass of ionized
hydrogen in the high-density, high-temperature limit, assuming that all 
nitrogen in the \Hii{} regions is singly ionized:
\begin{equation}\label{eq_mhii}
M_{\rm min}({\rm H\,{\sc II}})=\frac{L({\rm [N\,{\sc II}]_{205 \mu{\rm m}}}) \, m_{\rm H}}{(g_1 /g) \, A_{10} \, h \nu_{10} \, \chi(N^+)} \approx \, 2.27 \, \frac{L({\rm [N\,{\sc II}]_{205 \mu{\rm m}}})}{{\rm L_\odot}}\, {\rm M_\odot},
\end{equation}
where $A_{10}$ is the Einstein $A$ coefficient of the 
${}^3$P$_1\rightarrow^3$P$_0$ transition of nitrogen ($2.08\times10^{-6}$ 
s$^{-1}$); $g_1$=3 is the statistical weight of the $J=1$ level; 
$g = \Sigma_i g_i \exp(-\Delta E_i/k_b T)$ is the partition function, with 
$\Delta E_i$ being the energy difference between the fundamental and the 
$i$th level, $k_b$ being the Boltzmann constant and $T$ being the gas 
temperature; $h$ is Planck's constant, $\nu_{10}$=1461.1318 GHz is the 
rest-frame frequency of the transition; $m_{\rm H}$ is the mass of an 
hydrogen atom, and $\chi$(N$^+$) is the N$^+$/H$^+$ abundance ratio. In our 
working assumption, $\chi$(N$^+$) = $\chi$(N) $\approx 9.3\times10^{-5}$ 
\citep{savage96}. This gives 
$M_{\rm min}$({\rm H\,{\sc II}})=$4.1\times10^{9} \, \mu^{-1}$ \Msun{}
and $6.4\times10^{9} \, \mu^{-1}$ \Msun{} for \apm{} and \mm{}, 
respectively. 
It is interesting to compare these numbers with the molecular
gas mass estimated from the \aco{} luminosity \citep{riechers09,lestrade11}:
the minimum mass of the ionized gas is only a tiny fraction (0.8\% and 2.9\%, 
respectively) of the molecular reservoir in the two targets. Alternatively, 
the actual N$^+$/H$^+$ abundance ratio may be significantly lower than the 
adopted value. This is likely to happen if these systems are metal-poor (but 
this scenario is ruled out by the bright CO/FIR luminosity ratios observed in 
our targets), or if the ISM is enshrouded in a hard radiation field (i.e., 
nitrogen is multiply ionized). This is likely happening in \apm{}, which 
hosts a quasar (while no obvious signature of nuclear activity is observed in 
\mm{}). Indeed, \citet{ferkinhoff10} estimated that an ionized gas mass of 
$\sim3\cdot10^9$ \Msun{} is present in \apm{}, based on the 
\Oiii$_{88 \mu{\rm m}}$ line emission. These observations are complementary
to ours, as N is likely to be multiply ionized in the regions were most of
the \Oiii{} emission takes place.

\subsection{\niim{} contribution to ISM cooling}

Finally, we evaluate the role of \niim{} in the ISM cooling by comparing the 
\niim{} luminosity to the one of the FIR continuum. The latter is taken 
from the Spectral Energy Distribution fits by \citet{weiss07} for \apm{} 
($L_{\rm FIR}=20\times10^{13} \mu^{-1}$ \Lsun{})\footnote{Here we consider
the total FIR luminosity. However, \citet{weiss07} modeled the FIR emission
in \apm{} with two components, `warm' and `cold'. If the former is
powered from the quasar instead of star formation, the FIR luminosity 
should be scaled to a 10\% of the adopted value.}, and by \citet{lestrade10}
for \mm{} ($L_{\rm FIR}=48\times10^{13} \mu^{-1}$ \Lsun{}). The observed 
\niim{}/FIR luminosity ratio is $9.0\times10^{-6}$ and $5.8\times10^{-6}$ in 
the two sources. In Figure \ref{fig_nii_fir} we compare these values with 
the available measurements in the literature. Since the majority of data 
refer to the \Nii{}$_{\rm 122 \mu m}$ transition, we converted our 
estimates assuming a \Nii{} 122-to-205 \um{} luminosity ratio of 5 
\citep[see, e.g.,][]{beirao10}. Our observations confirm and extend the 
decreasing trend of the \Nii{}/FIR ratio as a function of the continuum 
luminosity, towards high-luminosities from local galaxies with
\Nii{}$_{\rm 122 \mu m}$/FIR$\sim$3$\times10^{-4}$
\citep{malhotra01,panuzzo10,fischer10,edge10,beirao10,vasta10,garcia11}  to FIR-luminous sources 
with \Nii{}$_{\rm 122 \mu m}$/FIR$\sim$3--5$\times10^{-5}$. Even in the 
extreme case of a 122-to-205 
\um{} ratio of 10, our data would populate a very different region 
of the plot with respect to the values found by \citet{ferkinhoff11}, who 
reported high \Nii{}$_{\rm 122 \mu m}$/FIR ratios ($\sim10^{-3}$) in two 
FIR-bright sources.

\section{Conclusions}

We present a study of the ionized ISM in two high-$z$ sources based on
the forbidden ionized nitrogen emission line at 205 \um{}. 
The FIR transitions of ionized nitrogen represents extremely powerful 
tools to study the 
properties of purely ionized gas in distant galaxies. A larger number of
\niim{} detections in distant galaxies is now mandatory in order to build 
a suitable sample for statistical analysis. This is now possible thanks
to the technological upgrades in the field. In particular, the 
unparalleled sensitivities reached by ALMA will open new possibilities
for the studies of high-$z$ objects.

\section*{Acknowledgments}
We thank the anonimous referee for his/her expert comments that increased
the quality of the manuscript. We thank B.\,Groves for useful discussions. 
This work is based on observations carried out with the IRAM Plateau de 
Bure Interferometer. IRAM is supported by INSU/CNRS (France), MPG (Germany)
and IGN (Spain). RD acknowledges funding from Germany's national research 
centre for aeronautics and space (DLR, project FKZ 50 OR 1104). DR 
acknowledges support from NASA through a Spitzer Space Telescope grant.

\label{lastpage}


\begin{thebibliography}{99}
\bibitem[\protect\citeauthoryear{Beir\~{a}o et al.}{2010}]{beirao10} Beir\~{a}o P., Armus L., Appleton P.N., Smith J.-D.T., Croxall K.V., Murphy E.J., Dale D.A., Helou G., et al., 2010, A\&A, 518, L60
\bibitem[\protect\citeauthoryear{Benford}{1999}]{benford99} Benford D.J., 1999, PhD Thesis, Caltech
\bibitem[\protect\citeauthoryear{Bigiel et al.}{2008}]{bigiel08} Bigiel F., Leroy A., Walter F., Brinks E., de Blok W.J.G., Madore B., Thornley M.D., 2008, AJ, 136, 2846
\bibitem[\protect\citeauthoryear{Bradford et al.}{2011}]{bradford11} Bradford C.M., Bolatto A.D., Maloney P.R., Aguirre J.E., Bock J.J., Glenn J., Kamenetzky J., Lupu R., Matsuhara H., Murphy E. J., et al. 2011, ApJ, 741, L37
\bibitem[\protect\citeauthoryear{Brown et al.}{1994}]{brown94} Brown J.M., Varberg T.D., Evenson K.M., \& Cooksy A.L., 1994, ApJ Letters, 428, 37
\bibitem[\protect\citeauthoryear{Colgan et al.}{1993}]{colgan93} Colgan S.W.J., Haas M.R., Erickson E.F., Rubin R.H., Simpson J.P., \& Russell, R.~W.\ 1993, ApJ, 413, 237
\bibitem[\protect\citeauthoryear{Combes et al.}{2012}]{combes12} Combes F., Rex M., Rawle T.D., Egami E., Boone F., Smail I., Richard J., Ivison R.J., et al., 2012, A\&A, in press (arXiv:1201.2908)
\bibitem[\protect\citeauthoryear{Cox et al.}{2011}]{cox11} Cox P., Krips M., Neri R., Omont A., G\"usten R., Menten K.M., Wyrowski F., Weiss A., et al., 2011, arXiv:1107.2924
\bibitem[\protect\citeauthoryear{Daddi et al.}{2010}]{daddi10} Daddi E., Elbaz D., Walter F., Bournaud F., Salmi F., Carilli C., Dannerbauer H., Dickinson M., et al., 2010, ApJ, 714, L118
\bibitem[\protect\citeauthoryear{De Breuck et al.}{2011}]{debreuck11} De Breuck C., Maiolino R., Caselli P., Coppin K., Hailey-Dunsheath S., Nagao T., 2011, A\&A, 530, L8
\bibitem[\protect\citeauthoryear{Downes et al.}{1999}]{downes99} Downes D., Neri R., Wiklind T., Wilner D.J., Shaver P.A., 1999, ApJ, 513, L1
\bibitem[\protect\citeauthoryear{Edge et al.}{2010}]{edge10} Edge A.C., Oonk J.B.R., Mittal R., Allen S.W., Baum S.A., B\"ohringer H., Bregman J.N., Bremer M.N., et al., A\&A, 518, L46
\bibitem[\protect\citeauthoryear{Egami et al.}{2000}]{egami00} Egami E., Neugebauer G., Soifer B.T., Matthews K., Ressler M., Becklin E.E., Murphy T.W.Jr., Dale D.A., 2000, ApJ, 535, 561
\bibitem[\protect\citeauthoryear{Ferkinhoff et al.}{2010}]{ferkinhoff10} Ferkinhoff C., Hailey-Dunsheath S., Nikola T., Parshley S.C., Stacey G.J., Benford D.J., Staguhn J.G., 2010, ApJ, 714, L147
\bibitem[\protect\citeauthoryear{Ferkinhoff et al.}{2011}]{ferkinhoff11} Ferkinhoff C., Brisbin D., Nikola T., Parshley S.C., Stacey G.J., Phillips T.G., Falgarone E., Benford D.J., Staguhn J.G., Tucker C.E., 2011, ApJ, 740, L29
\bibitem[\protect\citeauthoryear{Fischer et al.}{2010}]{fischer10} Fischer J., Sturm E., Gonz\'alez-Alfonso E., Graci\'a-Carpio J., Hailey-Dunsheath S., Poglitsch A., Contursi A., Lutz D., et al., A\&A, 518, L41
\bibitem[\protect\citeauthoryear{Garci\'a-Carpio et al.}{2011}]{garcia11} Garci\'a-Carpio J., Sturm E., Hailey-Densheath S., Fischer J., Contursi A., Poglitsch A., Genzel R., Gonz\'alez-Alfonso E., et al., 2011, ApJ, 728, L7
\bibitem[\protect\citeauthoryear{Genzel et al.}{2011}]{genzel10} Genzel R., Tacconi L.J., Gracia-Carpio J., Sternberg A., Cooper M.C., Shapiro K., Bolatto A., Bouch\'e N., et al., 2010, MNRAS, 407, 2091
\bibitem[\protect\citeauthoryear{Iono et al.}{2006}]{iono06} Iono D., Yun M.S., Elvis M., Peck A.B., Ho P.T.P., Wilner D.J., Hunter T.R., Matsushita S., Muller S., 2006, ApJ, 645, L97
\bibitem[\protect\citeauthoryear{Ivison \& Harrison}{1996}]{ivison96} Ivison R.J., Harrison A.P., 1996, A\&A, 309, 416
\bibitem[\protect\citeauthoryear{Ivison et al.}{2010}]{ivison10} Ivison R.J., Swinbank A.M., Swinyard B., Smail I., Pearson C.P., Rigopoulou D., Polehampton E., Baluteau J.-P., et al., 2010, A\&A, 518, L35
\bibitem[\protect\citeauthoryear{Kennicutt}{1998}]{kennicutt98} Kennicutt R.C., 1998, ApJ, 498, 541
\bibitem[\protect\citeauthoryear{Krips et al.}{2007}]{krips07} Krips M., Peck A.B., Sakamoto K., Petitpas G.B., Wilner D.J., Matsushita S., Iono D., 2007, ApJ, 671, L5
\bibitem[\protect\citeauthoryear{Leroy et al.}{2008}]{leroy08} Leroy A.K., Walter F., Brinks E., Bigiel F., de Blok W.J.G., Madore B., Thornley M.D., 2008, AJ, 136, 2782
\bibitem[\protect\citeauthoryear{Lestrade et al.}{2010}]{lestrade10} Lestrade J.-F., Combes F., Salom\'e P., Omont A., Bertoldi F., Andr\'e P., Schneider N., 2010, A\&A, 522, L4
\bibitem[\protect\citeauthoryear{Lestrade et al.}{2011}]{lestrade11} Lestrade J.-F., Carilli C.L., Thanjavur K., Kneib J.-P., Riechers D.A., Bertoldi F., Walter F., Omont A., 2011, ApJ, 739, L30
\bibitem[\protect\citeauthoryear{Loenen et al.}{2010}]{loenen10} Loenen A.F., van der Werf P.P., G\"usten R., Meijerink R., Israel F.P., Requena-Torres M.A., Garc\'ia-Burillo S., Harris A.I.,et al., 2010, A\&A, 521, L2
\bibitem[\protect\citeauthoryear{Luhman et al.}{1998}]{luhman98} Luhman M.L., et al., 1998, ApJ Letters, 504, 11
\bibitem[\protect\citeauthoryear{Madden et al.}{1997}]{madden97} Madden S.C., Poglitsch A., Geis N., Stacey G.J., \& Townes C.H., 1997, ApJ, 483, 200
\bibitem[\protect\citeauthoryear{Maiolino et al.}{2005}]{maiolino05} Maiolino R., Cox P., Caselli P., Beelen A., Bertoldi F., Carilli C.L., Kaufman M.J., Menten K.M., Nagao T., Omont A., et al., 2005, A\&A, 440, L51
\bibitem[\protect\citeauthoryear{Maiolino et al.}{2009}]{maiolino09} Maiolino R., Caselli P., Nagao T., Walmsley M., De Breuck C., Meneghetti M., 2009, A\&A, 500, L1
\bibitem[\protect\citeauthoryear{Malhotra et al.}{2001}]{malhotra01} Malhotra S., Kaufman M.J., Hollenbach D., Helou G., Rubin R.H., Brauher J., Dale D., Lu N.Y., et al., 2001 ApJ, 561, 766
\bibitem[\protect\citeauthoryear{Matteucci \& Padovani}{1993}]{matteucci93} Matteucci F., Padovani P., 1993, ApJ, 419, 485
\bibitem[\protect\citeauthoryear{Oberst et al.}{2006}]{oberst06} Oberst T.E., et al.\ 2006, ApJ Letters, 652, 125
\bibitem[\protect\citeauthoryear{Osterbrock \& Ferland}{2006}]{osterbrock06} Osterbrock D.E. \& Ferland G.J., 2006, Astrophysics of gaseous nebulae and active galactic nuclei
\bibitem[\protect\citeauthoryear{Panuzzo et al.}{2010}]{panuzzo10} Panuzzo P., Rangwala N., Rykala A., Isaak K.G., Glenn J., Wilson C.D., Auld R., Baes M., et al., 2010, A\&A, 518, L37
\bibitem[\protect\citeauthoryear{Petuchowski \& Bennett}{1993}]{petuchowski93} Petuchowski S.J., \& Bennett C.L., 1993, ApJ, 405, 591
\bibitem[\protect\citeauthoryear{Riechers et al.}{2009}]{riechers09} Riechers D.A., Walter F., Carilli C.L., Lewis G.F., 2009, ApJ, 690, 463
\bibitem[\protect\citeauthoryear{Savage \& Sembach}{1996}]{savage96} Savage B.D., \& Sembach K.R., 1996, ARA\&A, 34, 279
\bibitem[\protect\citeauthoryear{Schruba et al.}{2011}]{schruba11} Schruba A., Leroy A.K., Walter F., Bigiel F., Brinks E., de Blok W.J.G., Dumas G., Kramer C., et al., 2011, AJ, 142, 37
\bibitem[\protect\citeauthoryear{Solomon et al.}{1992}]{solomon92} Solomon P.M., Downes D., \& Radford S.J.E., 1992, ApJ, 398, L29
\bibitem[\protect\citeauthoryear{Stacey et al.}{1991}]{stacey91} Stacey G.J., Geis N., Genzel R., Lugten J.B., Poglitsch A., Sternberg A., \& Townes C.H., 1991, ApJ, 373, 423
\bibitem[\protect\citeauthoryear{Stacey et al.}{2010}]{stacey10} Stacey G.J., Hailey-Dunsheath S., Ferkinhoff C., Nikola T., Parshley S.C., Benford D.J., Staguhn J.G., Fiolet N., 2010, ApJ, 724, 957
\bibitem[\protect\citeauthoryear{Stasinska}{2007}]{stasinska07} Stasinska G., 2007, arXiv:0704.0348
\bibitem[\protect\citeauthoryear{van der Werf}{1999}]{vanderwerf99} van der Werf P.P., 1999, Highly Redshifted Radio Lines, 156, 91
\bibitem[\protect\citeauthoryear{van der Werf}{2010}]{vanderwerf10} van der Werf P.P., Isaak K.G., Meijerink R., Spaans M., Rykala A., Fulton T., Loenen A.F., Walter F., Wei\ss{} A., Armus L., et al., 2010, A\&A, 518, L42
\bibitem[\protect\citeauthoryear{Vasta et al.}{2010}]{vasta10} Vasta M., Barlow M.J., Viti S., Yates J.A., Bell T.A., 2010, MNRAS, 404, 1910
\bibitem[\protect\citeauthoryear{Wagg et al.}{2006}]{wagg06} Wagg J., Wilner D.J., Neri R., Downes D., Wiklind T., 2006, ApJ, 651, 46
\bibitem[\protect\citeauthoryear{Wagg et al.}{2010}]{wagg10} Wagg J., Carilli C.L., Wilner D.J., Cox P., De Breuck C., Menten K., Riechers D.A., Walter F., 2010, A\&A, 519, L1
\bibitem[\protect\citeauthoryear{Walter et al.}{2009a}]{walter09a} Walter F., Weiss A., Riechers D.A., Carilli C.L., Bertoldi F., Cox P., Menten K.M., 2009a, ApJ, 691, L1
\bibitem[\protect\citeauthoryear{Walter et al.}{2009b}]{walter09b} Walter F., Riechers D., Cox P., Neri R., Carilli C., Bertoldi F., Weiss A., Maiolino R., 2009b, Nature, 457, 699
\bibitem[\protect\citeauthoryear{Walter et al.}{2011}]{walter11b} Walter F., Weiss A., Downes D., Decarli R., Henkel C., 2011, ApJ, 730, 18
\bibitem[\protect\citeauthoryear{Wei\ss{} et al.}{2007}]{weiss07} Wei\ss{} A., Downes D., Neri R., Walter F., Henkel C., Wilner D.J., Wagg J., Wiklind T., 2007, A\&A, 467, 955
\bibitem[\protect\citeauthoryear{White et al.}{2010}]{white10} White G.J., Abergel A., Spencer L., Schneider N., Naylor D.A., Anderson L.D., Joblin C., Ade P., Andr\'e P., Arab H., et al., 2010, A\&A, 518, L114
\bibitem[\protect\citeauthoryear{Wong \& Blitz}{2002}]{wong02} Wong T., \& Blitz L., 2002, ApJ, 569, 157
\bibitem[\protect\citeauthoryear{Wright et al.}{1991}]{wright91} Wright E.L., et al., 1991, ApJ, 381, 200
\end{thebibliography}
\end{document}